\documentclass[preprint,12pt]{elsarticle}

\usepackage{amssymb}
\usepackage{amsmath}
\usepackage{longtable}
\usepackage{subcaption}
\usepackage{multirow}
\usepackage{footnote}
\usepackage{caption}
\makesavenoteenv{longtable}

\journal{International Journal of Forecasting}

\begin{document}

\begin{frontmatter}



\title{Monthly GDP nowcasting with Machine Learning and
Unstructured Data}


\author[inst1]{Juan Tenorio}
\affiliation[inst1]{organization={Universidad Peruana de Ciencias Aplicadas},
            addressline={2390 Prolongación Primavera}, 
            city={Santiago de Surco},
            postcode={15023}, 
            state={Lima},
            country={Peru, pcefjten@upc.edu.pe}}

\author[inst2]{Wilder Perez}
\affiliation[inst2]{organization={Universidad Científica del Sur},
            addressline={Panamericana Sur Km 19}, 
            city={Chorrillos},
            postcode={15067}, 
            state={Lima},
            country={Peru}}

\begin{abstract}
In the dynamic landscape of continuous change, Machine Learning (ML) ``nowcasting" models offer a distinct advantage for informed decision-making in both public and private sectors. This study introduces ML-based GDP growth projection models for monthly rates in Peru, integrating structured macroeconomic indicators with high-frequency unstructured sentiment variables. Analyzing data from January 2007 to May 2023, encompassing 91 leading economic indicators, the study evaluates six ML algorithms to identify optimal predictors. Findings highlight the superior predictive capability of ML models using unstructured data, particularly Gradient Boosting Machine, LASSO, and Elastic Net, exhibiting a 20\% to 25\% reduction in prediction errors compared to traditional AR and Dynamic Factor Models (DFM). This enhanced performance is attributed to better handling of data of ML models in high-uncertainty periods, such as economic crises.
\end{abstract}




\end{frontmatter}


\section{Introduction}
\label{sec:introduction}
Making decisions in real-time is a true challenge for policymakers, given that the primary barrier they face is the usual delay in the availability of updated information about macroeconomic aggregates. In most cases, the economic variables show a delay of between 30-45 days on average, including the time for revisions and retrospectives. To address the issue of extended delays in the publication of key economic aggregates, the concept of nowcasting is proposed, which aims to predict the present, the very near future and the very recent past \cite{giannone2013now}. One of the most traditional nowcasting approaches is the Dynamic Factor Model (DFM) which is a widely used method in central banks to predict GDP {\cite{giannone2008nowcasting, banbura2011look, stock1989new, longo2022neural,bok2018macroeconomic,rusnak2016nowcasting,gonzalez2019nowcasting}}. For instance, \cite{giannone2008nowcasting} proposed a methodology to assess the marginal impact of the publication of monthly-updated data on forecasts of quarterly-published real Gross Domestic Product (GDP) growth. The method presented by these authors was able to track the real-time flow of information that central banks monitor by handling large datasets with staggered publication dates and updating primary forecasts each time new higher-frequency data is published. Another seminal study was proposed by \cite{evans2005we} where they do real-time estimations of the current state of the US economy. This approach included data complexity and provided useful information about the relationship between macroeconomics and asset prices.

A critical challenge to that traditional approach is the increase in uncertainty in the estimates which use a limited set of variables, and often fall short. Nevertheless, the continuous stride forward in the new generation of high-frequency data has changed how prediction models face the uncertainty inherent in this information. As a result, in the recent few years, both central banks and international institutions have adopted methodological focuses that incorporate machine learning, and take advantage of the abundant quantity of data that come from search engines and social media such as \cite{richardson2018nowcasting}, \cite{zhang2023nowcasting} and {\cite{gonzalez2019nowcasting}. Those methods provide more accurate predictions by incorporating various variables and new sources of unstructured data. As described {\cite{athey2018impact}, these techniques are divided into two main brands, supervised and unsupervised ML. {\cite{athey2018impact} explains that unsupervised MLs are looking for groups of observations that are similar in terms of their covariance. Thus, a ``dimensionality reduction" can be performed. Unsupervised MLs commonly use videos, images, and text as a source of information, in techniques such as grouping \textit{k-medias}.  For instance, \cite{blei2003latent} applied pooling models to find ``topics" in textual data. Another example is the paper written by \cite{woloszko2020weekly}. Here, the author shows a weekly indicator of the economic activity for 46 OCDE countries and the G20 using search data from Google Trends. This document showcases the power of prediction of specific ``topics", including ``bankruptcies", ``economic crises", ``investment", ``baggage" and ``mortgages". Calibration is performed using a neural network that captures nonlinear patterns, which are shown to be consistent with economic intuition using ML Shapley values interpretation tools. On the other side, the supervised ML algorithms as pointed out by \cite{varian2014machine} imply the use of a group of variables features o co variables to predict a specific indicator result. There is a variety of supervised ML methods regressions such as  \textit{LASSO}, \textit{Ridge}, \textit{Elastic Net},\textit{Random Forest}, \textit{Regression Trees}, \textit{Support Vector Machines}, \textit{Neural Nets}, \textit{Matrix Factorisation}, among others as \textit{Model Averaging}.

Several studies highlight the advantages of supervised ML models to forecast macroeconomic series that overcome traditional methods. An application is the research of \cite{ghosh2023machine}, who present a compilation of Machine Learning techniques and conventional time series methods to predict the Indian GDP. They estimate the ML in the DFM context with financial and economic uncertainty data. They estimate machine learning models such as Random Forest and Prophet along with conventional time series models such as ARIMA to nowcast Indian GDP, where hybrid models stand out. Likewise, the results from \cite{richardson2018nowcasting} showed better performance of the Ridge regression model to the nowcast GDP of New Zealand over a Dynamic Factor Model. \cite{muchisha2021nowcasting} built and compared ML models to forecast the GDP of Indonesia. They evaluate six ML algorithms: Random Forest, LASSO, Ridge, Elastic Net, Neural Networks and Support Vector Machines. Their results make clear the outstanding performance of ML than auto-regressive models, especially the Random Forest model. Also, \cite{zhang2023nowcasting} test ML, DFM and static factor and MIDAS regressions models to nowcast the GDP rate growth of China. They find superior accuracy of ML compared to DFM. The ML model that deserves more attention was Ridge Regression, which overcame the others not only on prediction but also early anticipation of crises such as the global financial crisis and COVID-19. \cite{kant2022nowcasting} compare models to the Netherlands economy between 1992 and 2018, where Random Forest algorithms stood out. \cite{suphaphiphat2022scalable} use novel variables such as Google Search and air quality. They run standard DFM and ML to European economies during normal times and crises. They show that most MLs significantly outperform the AR(1) reference model. They highlight that DFM tends to perform better in normal times, while many of the ML methods have excellent performance in identifying turning points.

Also, te recent literature has highlighted the relevance of incorporating Big Data due to its benefits of broadening the range and use of available data that can provide some valid information on the behaviour of the economy to anticipate certain economic indicators in real time \cite{einav2014data}. As mentioned in \cite{eberendu2016unstructured}, the digital era has allowed the emergence of news channels and social network technologies, mobile phones and online advertising, which are a new kind of source data, but without a pre-fixed format raises new challenges. This data is available in formats like text, XML, email, images, videos, etc. In that sense, we can denominate these data as non-structured. \cite{eberendu2016unstructured} gives and general description of this type of data. A seminal example is the study of \cite{varian2014machine} which indicates how the search related to the “initial claims for unemployment” in Google Trends are good candidates to forecast unemployment, CPI and consumer confidence in countries such as the US, UK, Canada, Germany and Japan. They focus on immediate out-of-sample forecasting and extend the Bayesian structural time series model using the Hamiltonian sampler for variable selection. These authors obtain good results for unemployment, while for CPI or consumer confidence not so good.

A characteristic of these algorithms that is often highlighted resides in their capacity to formulate parametric selections in big data sets, which find their base in training a specific percentage of the model’s information. Hence, the objective of this document consists of exploring the benefits of using diverse machine learning methodologies. This will be done by combining the use of conventional leading indicators (structured data) and indicators of analysis of sentiment (non-structured data) to build a precise monthly indicator of Peru’s real GDP (Gross Domestic Product) growth. The data set consists of both local and international variables, which can be broken down into 53 structured variables in 38 non-structured variables, giving a total of 91 predictors. These predictive variables are examined according to the model, to evaluate the optimum performance of each variable between September 2014 and May 2023. Furthermore, following \cite{romer2008fomc}, an analysis of the evaluation of prediction accuracy will be performed using two models as benchmarks: the traditional autoregressive time series, and three specifications of dynamic factor models, also we assess the robustness of the results by building an additional DFM incorporating only the indicator of electricity commonly used by policymakers and consulting firms to track the economic activity, especially in Peru. This will facilitate an exhaustive evaluation of the performance of machine learning algorithms.

The results indicate that immediate predictions of the machine learning models are more solid in comparison with the benchmark auto-regressive model and DFM models. Specifically, the Random Forest, Gradient Boosting Machine, and Adaptive Lasso show performance with a superior ability to reduce the average error of projection in a range from 20\%-25\%. Additionally, it is corroborated that following the methodology proposed by \cite{armstrong2001principles}, the utilization of the average value of projection of all the machine learning algorithms, adds a significant value to the RMSE, which positively contributes to a more precise prediction of GDP. Even though other methodologies, such as \textit{Ridge}, \textit{LASSO} and \textit{Elastic Net} do not reach the same level of predictive ability as the previously mentioned methodologies, they still outperform the control model in terms of performance. Further, the proof of predicting evaluation and exercises of consistency, confirm that most of the machine learning models improve the prediction significantly which falls in line with previous literature applied in other contexts \cite{richardson2018nowcasting, varian2014machine, zhang2023nowcasting}.

This research document adds itself to the existing literature that highlights the success of machine learning applications in contrast to more traditional methodologies. However, given the lack of evidence in Latin America, and in particular in Peru\footnote{See \cite{escobal2002sistema}} who built a joint leading indicator to tracking the Peruvian GDP with only 14 variables. \cite{kapsoli2002indicadores} perform forward GDP estimation with a nonlinear neural network model. \cite{etter2011composite} proposed a leading indicator with the expectations survey. \cite{martinez2014indicador} estimated the growth rate of GDP based on electric production. Following to \cite{aruoba2009real}, \cite{forero2016indicador} and \cite{perez2018nowcasting} proposed a leading indicator to Peruvian GDP.}, surrounding the use of these algorithms in conjunction with non-structured data, this research project also highlights the need to bring to the forefront of the discussion what these models entail. Previous works applied to the Peruvian economy are focused on the anticipated estimation of monthly GDP growth based on a set of leading indicators (structured data). However, a scarce application of machine learning models and the inclusion of unstructured data in GDP forecasting is evident. In the Latin America case we found the work of \cite{barrios2021nowcasting}, \cite{richardson2018nowcasting} and \cite{dopke2017predicting} who have shown through the implementation of diverse machine learning algorithms that these method’s results are more adequate in carrying out forecasts in real-time when a large amount of information is at the researcher’s disposal. Similarly, in the case of El Salvador and Belize, \cite{barrios2021nowcasting} implemented a large array of machine, learning methods to forecast the quarterly growth of GDP, using a large amount of predictive variables. The results of this research study concluded that the application of these tools represents a solid alternative to prediction, and its benefits suggest a recommendation for its use in other countries in the region. Additionally, other researchers have extended the application of the machine learning models to GDP, inflation, yield curve, and active prices. These efforts have yielded notable results in precise forecasting \cite{medeiros2021forecasting, giglio2022factor}.

It is still important to highlight that these methods present challenges in their implementation, which have led to some major debates surrounding the topic. In fact, \cite{green2015simple}, as well as \cite{makridakis2018statistical}, when comparing multiple models of machine learning, find that the deposition of the forecasting is less significant in comparison with the statistical smoothing approaches, and the ARIMA models. These authors warn that the computational complexity that is inherent to the selection and use of variables in the machine learning model makes immediate forecasting difficult and less practical for policymakers.

Finally, the rest of this document is structured in several sections. Initially, a section focused on the methodology is presented in which details are given about the ML models and traditional models that were used and the data (structured and non-structured) sets. Afterwards, the results of the main estimations are displayed in a specific section \ref{sec:results}, followed by the sections of robustness assessment and the conclusion.

\section{Methodology}
\label{sec:meth}
This section provides a brief description of the different regularization methods and decision trees used to select the best predictors for the monthly nowcasting model and to calibrate the hyperparameters, in a series from January 2007 to May 2023. The six methods that are used are Random Forest (RF), Gradient Boosting Machine (GBM), LASSO regression, Ridge, Elastic Net, and as a benchmark, an autoregressive (AR) and dynamic factor model (DFM) are utilized.

\subsection{Autoregressive Model (AR)}
As a starting point for our reference, we establish an autoregressive AR model for the monthly GDP growth ($y_t$), which reflects the value of a variable in terms of its previous values. A model of order 1, following these characteristics, exhibits the following structure:

\begin{equation}
    y_t = \beta_0 + \beta_1 y_{t-1}+e_t  
\end{equation}

where $\beta_0$ is a constant term, $\beta_1$ is a parameter, and $e_t$ is a term that represents the error and captures the randomness of the model. 

\subsection{Dynamic Factor Model (DFM)}
DFMs are estimated in the form of state-space systems and can be estimated using the Kalman filter and various types of algorithms. The one of the most popular in the economic literature is the Expectation Maximization algorithm, due to its robust numerical properties following the proposal by \cite{doz2011two}, which is an efficient estimation to bigger datasets.

The canonical reference DFM can be described as follows: 

\begin{align}
x_t = C_0 f_{t} + e_{t} \ \ \ \ \  & e_t \sim N(0,R) \\
f_t = \sum_{j=1}^{p} A_j f_{t-j} + u_{t} \ \ \ \ \  & u_t \sim N(0,Q_0)
\end{align}

Where equation 2 is identified as the measurement equation and equation 3 as the transition equation, allowing the unobservable factor $f_t$ to evolve as in a vector autoregressive model. These equations do not include trends or intercepts, as the included data must be stationary and standardized before estimation.

The matrix system is as follows:

$x_t$: a vector of $n\times 1$ observable time series at time $t : (x_t, ..., x_{nt})'$, which allows for missing data.\

$f_t$: a vector of $r\times 1$ factors at time $t : (f_t, ..., f_{rt})'$.\

$C_0$: a matrix of $n\times r$ observable time series with lag $j$.\

$Q_0$: a matrix of $r\times r$ state covariances.\

$R$: a matrix of $r\times r$ measurement covariances. This matrix is diagonal under the assumption that all covariances between the series are explained by the factors $ E [x_{it} \mid x_{-i,t}, f_t] = c_{0i} f_{t} \forall i $, where $c_{0i}$ is the $i-th$ row of $C_0$.

This model can be estimated using a classical form of the Kalman Filter and the Maximum Likelihood estimation algorithm, after transforming it into a State Space model. In a VAR expression, it would be as follows:
\begin{align}
    x_t = C F_{t} + e_{t}    \ \ \ \ \  & e_t \sim N(0,R) \\
    F_t = A F_{t-1} + u_{t} \ \ \ \ \  & u_t \sim N(0,Q) 
\end{align}

As a benchmark model, we use the efficient estimation of a Dynamic Factor Model via the EM Algorithm - on stationary data with time-invariant system matrices and classical assumptions while permitting missing data following the approach of \cite{banbura2014maximum}.

\subsection{Penalized Regression Models}
These methodologies are employed to optimize the selection of predictor variables and control the model's complexity, which is crucial in preventing overfitting in high-dimensional settings. The literature suggests various forms of penalization to estimate the parameters $\beta_j$ accurately. We will briefly explore the characteristics of the Ridge, Lasso, Elastic Net, and Adaptive Lasso models, emphasizing how these techniques allow for proper weighting of coefficients and how their application impacts the inclusion and relevance of variables in the final model.

\subsubsection{Ridge Regression}
The Ridge model is defined by adding a penalty based on the sum of squares of the coefficients of the predictor variables. This penalty compels the coefficients to be very small, preventing them from taking extremely high values, thus reducing the influence of less relevant variables. To estimate the coefficients $\hat{\beta}^{Ridge}$, the equation must be expressed as:

\begin{equation}
    \min_{\beta} \left( \sum_{i=1}^{n} (y_i - \beta_0 - \sum_{j=1}^{p} x_{ij}\beta_j)^2 + \lambda \sum_{j=1}^{p} \beta_j^2 \right)  
\end{equation}

Where $y_i$ is the observed value of the dependent variable for observation $i$, $x_{ij}$ is the value of predictor variable $j$ in observation $i$, $\beta_j$ is the coefficient associated with predictor variable $j$, $p$ is the number of predictor variables, and $\lambda$ is the regularization hyperparameter that controls the magnitude of the penalty. The sum of the terms $\beta_j^2$ in the penalty prevents the coefficients from reaching large values, thereby contributing to stability and reducing the risk of overfitting.

\subsubsection{LASSO Regression}
The LASSO (Least Absolute Shrinkage and Selection Operator) model, introduced by \cite{tibshirani1996regression}, employs a penalty based on the sum of the absolute values of the coefficients of the predictor variables. This penalty has the property of forcing some coefficients to exactly reach zero, resulting in the automatic selection of a subset of more relevant predictor variables and the elimination of less significant ones. The Lasso coefficients $\hat{\beta}^{Lasso}$ are estimated:

\begin{equation}
 \min_{\beta} \left( \sum_{i=1}^{n} (y_i - \beta_0 - \sum_{j=1}^{p} x_{ij}\beta_j)^2 + \lambda \sum_{j=1}^{p} |\beta_j| \right)
\end{equation}

The change lies in the hyperparameter $\lambda$ which, by summing the absolute values of the coefficients $|\beta_j|$ in the penalty, leads to model selection and simplification by allowing some coefficients to be zero. This provides a more precise variable selection approach regarding the degree of importance of all variables.

\subsubsection{Elastic Net Regression}
The Elastic Net model appropriately combines the constraints of both the LASSO and Ridge models. In particular, \cite{zou2005regularization} mention that its advantage lies in correcting the model when the number of regressors exceeds the number of observations ($p>n$), which improves variable grouping. The penalty includes both the sum of the absolute values of the coefficients and the sum of the squares of the coefficients of the predictor variables. The equation for estimating the coefficients $\hat{\beta}^{Enet}$ is expressed as:
\begin{equation}
 \min_{\beta} \left( \sum_{i=1}^{n} (y_i - \beta_0 - \sum_{j=1}^{p} x_{ij}\beta_j)^2 + \lambda \sum_{j=1}^{p} \left( \alpha |\beta_j| + (1 - \alpha)  \beta_j^2 \right) \right)
\end{equation}

where \(\lambda\)  is the global regularization hyperparameter and \(\alpha\) is the hyperparameter that controls the mix between Lasso (\(\alpha = 1\)) and Ridge (\(\alpha = 0\)) penalties. The combination of both penalties in the Elastic Net model allows for a higher degree of flexibility in variable selection and coefficient alignment.

\subsubsection{Adaptive Lasso Regression}
Following \cite{zou2006adaptive}, the Adaptive LASSO model is a variant of the LASSO model that introduces a penalty approach which adaptively adjusts the magnitude of the penalties for each coefficient of the predictor variables. This adaptation allows for penalties to be different for different coefficients, potentially resulting in a more precise selection of relevant variables. \cite{liu2014doubly},  indicate that this process can be efficiently performed using the LARS algorithm. The equation for the Adaptive LASSO model ($\hat{\beta}^{AdL}$) is expressed as:

\begin{equation}
 \min_{\beta} \left( \sum_{i=1}^{n} (y_i - \beta_0 - \sum_{j=1}^{p} x_{ij}\beta_j)^2 + \lambda \sum_{j=1}^{p} w_j |\beta_j| \right)
\end{equation}

where \(\lambda\) is the regularization hyperparameter, and $w_j$ is the adaptation factor for the coefficient $\beta_j$. It is important to note that the exact form of the adaptation factors \(w_j\) depends on the specific implementation and may vary. In general, these factors are calculated based on the absolute values of the coefficients in previous iterations of the algorithm.

\subsection{Decision Tree Models}
Decision Tree models are machine learning algorithms that represent decisions and actions in the form of a tree. In this case, we will present two algorithms where each internal node of the tree represents a feature or attribute, and each branch represents a decision or rule based on that attribute. The training data is divided based on these decisions until it reaches leaf nodes, which correspond to the predictions, in our case, related to monthly GDP growth. Additionally, the use of these trees allows for an improvement in variable selection by handling non-linear relationships in the model.

\subsubsection{Random forest}
This method is based on constructing decision trees using variables from a matrix $X$ and a random selection of features. Additionally, it involves randomly selecting subsets of data from $X$ with replacement to train each tree in the ensemble, distinguishing it from other tree-based techniques. Each tree generates a prediction of the target variable (in this case, monthly GDP), and the final model selects the most voted prediction in the ensemble of trees \cite{breiman2001random}. According to \cite{tiffin2016seeing}, Random Forest has the advantage of combining predictions from multiple trees and selecting those with lower error, thereby reducing the influence of potential individual errors (if the correlation between trees is low). In summary, this method recursively divides the data in $X$ into optimized regions and uses variable-based criteria to forecast the target variable, then calculates the dependent variable as the average of these regions.

\begin{equation}
   \hat{f}(x)=\sum_{m}\hat{c}_m I(x \in X_r); \hat{c}_m=avg(y_i|x_i \in X_r) 
\end{equation}

The algorithm has certain advantages, such as being efficient in handling large datasets with many variables, providing an estimation of variable importance, and offering an unbiased estimation of generalization error during its construction \cite{breiman2001random}. However, it has disadvantages like difficulty in interpreting results beyond predictions and a computationally intensive demand for training and hyperparameter tuning. Therefore, for this model, it was necessary to fine-tune it through cross-validation, achieving better performance on unseen data. 

\begin{figure}[!ht]
\centering
\begin{minipage}[t]{.9\linewidth}\label{fig:1}
\centering
\caption{{Simple Representation of the Random Forest Algorithm}}
 \includegraphics[width=.9\linewidth]{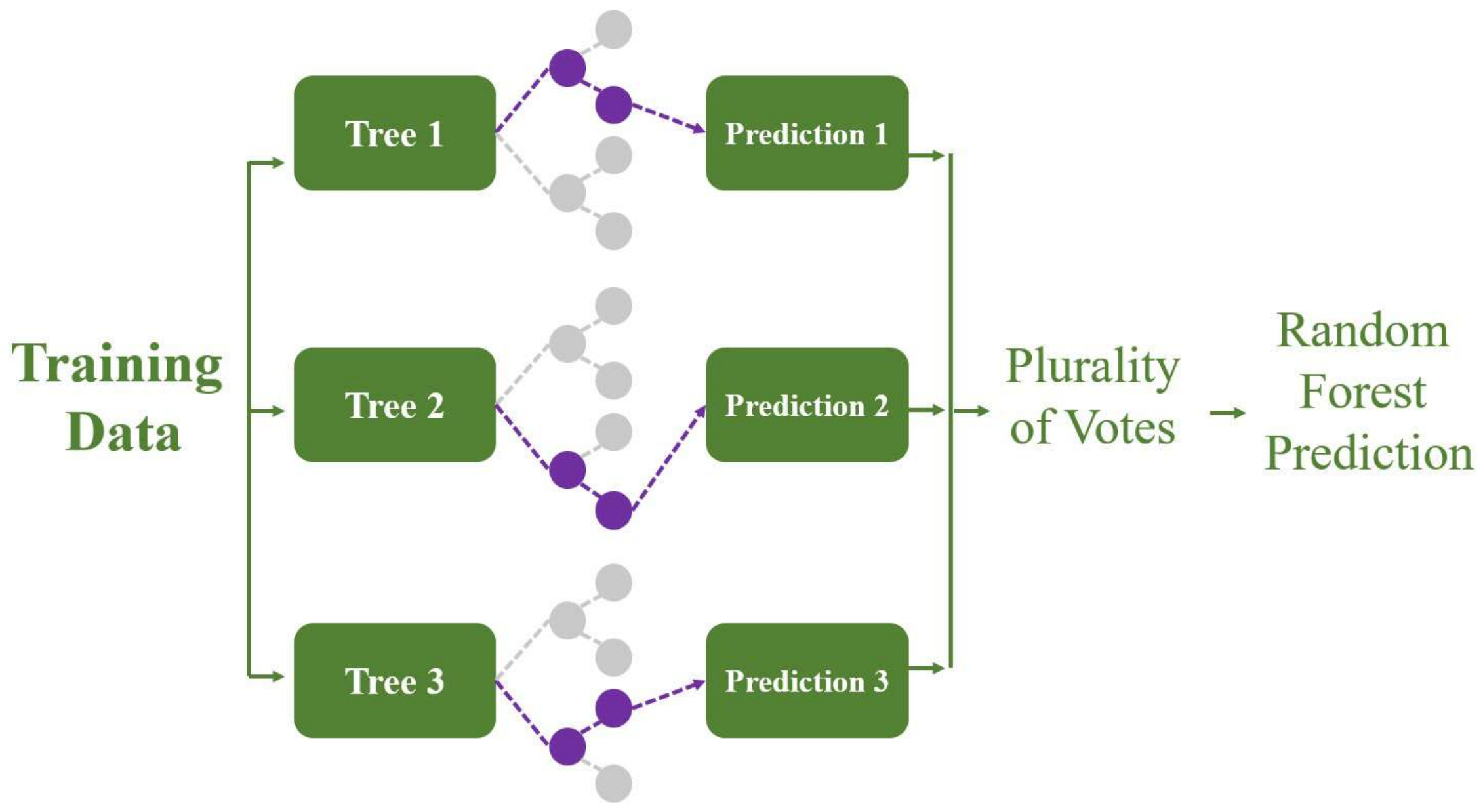}
\end{minipage}\\
\raggedright
\footnotesize{Source: Own elaboration}
\end{figure}

\subsubsection{Gradient Boosting Machine}
"This algorithm builds a sequence of decision trees, where each tree is fitted to the residual errors of the previous tree. Therefore, each iteration obtains a new tree that minimizes the remaining error. These prediction models are trained using the errors from the accumulated set of weak predictions \cite{brownlee2016bagging}, indicates that weak models do not necessarily mean they are better than accurate models, as they have the advantage of being able to correct the overfitting problem.} in a way that provides a progressive improvement in regression performance compared to the initial model \cite{natekin2013gradient}.

\begin{figure}[h]
\centering
\begin{minipage}[t]{.9\linewidth}\label{fig:2}
\centering
\caption{{Simple Representation of the Gradient Boosting Machine Algorithm}}
 \includegraphics[width=.9\linewidth]{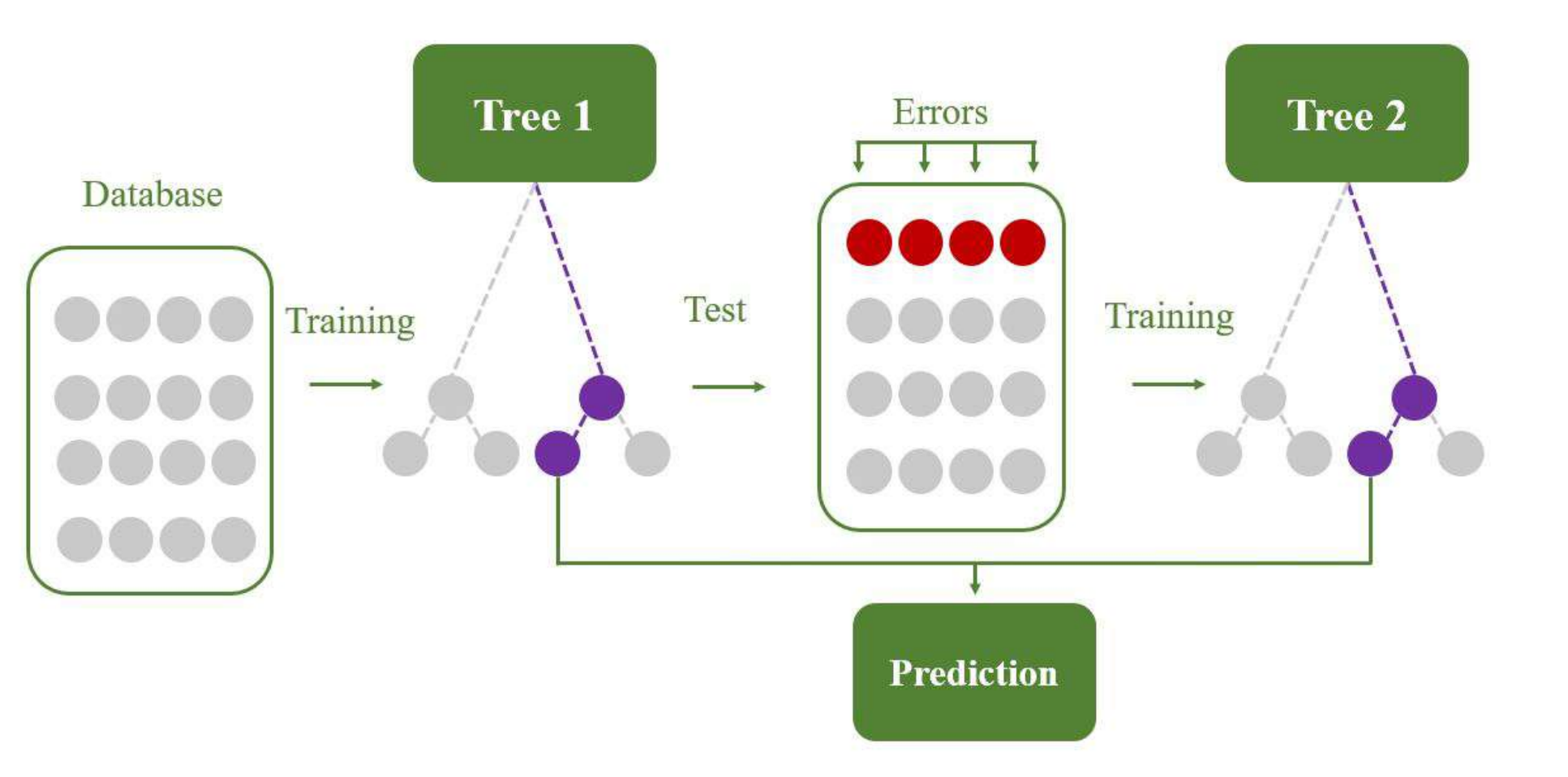}
\end{minipage}\\
\raggedright
\footnotesize{Source: According to \cite{boehmke2020chapter}}
\end{figure}

In essence, each tree in this algorithm contributes its prediction, which is added to the sequence of predictions from previous trees in order to enhance the final prediction of the model. \cite{boehmke2020chapter}, mention that this method can be summarized by the following equation.

\begin{equation}
   F(x) = \sum^{Z}_{z=1}F_z(x) 
\end{equation}

where $z$ is the number of trees that cumulatively add up the errors from all preceding trees. That is, the first tree $y=F_1(x)$, then the second tree will be $F_2(x)=F_1(x)+e_1$ and so on, successively, to minimize $F(x)$ as the following expression:

\begin{equation}
   L= \sum_{z} L(y_z,F_z(x)) 
\end{equation}

Therefore, as new decision trees are incorporated, the accuracy of the final projection improves gradually, resulting in more precise forecasts for monthly GDP.

\subsection{Data}
The model's database comprises a variety of variables, ranging from macroeconomic and financial data to unstructured information related to sentiment or “trend" (See Tables 6, 7, and 8). This information set encompasses consumption indicators, such as credits, deposits, chicken sales, consumer surveys, and local activity indicators, including electricity production, hydrocarbons, economic expectations, and others. Investment indicators are also incorporated, such as internal cement consumption, capital goods imports, and so forth. A set of monetary indicators covering consumer and producer price indices, among others, is included. It is important to highlight the inclusion of sectoral variables related to fishing and agricultural production, which constitutes a unique feature compared to other nowcasting models. Furthermore, the database covers information on foreign trade, the labour market, and climate data.

In addition to conventional variables, we have incorporated unstructured data related to perception in various areas, such as the economy, consumption, labour market, politics, tourism, government support, and natural phenomena. These variables can capture the general sentiment of the population and its potential influence on economic indicators. In particular, the use of massive search engines, such as Google, stands out as a powerful tool for providing real-time information. \cite{scott2015bayesian}, have pointed out that the inclusion of online searches as variables provides substantial benefits to short-term forecasting models, especially in detecting periods of high volatility. This is demonstrated in the ability to anticipate both the recession caused by the COVID-19 pandemic and the subsequent period of economic recovery. Consequently, the effectiveness of this approach has been widely investigated and adopted by central banks and international institutions. Thus, we estimate 10 groups (See Table 6) of variables that aim to track Google search queries, which are updated daily and can be downloaded from Google Trends. The selection of these words (variables) aims to convey different aspects of the economy, such as the consumption-related group, which is constructed based on searches for words like “Kia", “Restaurants", “Toyota", “Credits", “Loans", “Deals", “Mortgages", and “Cinema". Once this textual data is converted into numerical data, the inclusion of these series is evaluated in the estimations of an optimal model using Gibbs sampling following the findings of \cite{garcia2013sampling} and using  50,000 iterations, an initial burning of 1,000 iterations, and constant beta priors (see Figure 10). This indicates that there is a high relevance of the group of unstructured variables such as the search frequency for “flights", “peruflight\_us", “visa", or “El Niño", which would reflect the dynamics of tourism and climatic conditions, among others. Furthermore, we compare the results of this estimation with another one by reducing the sample to 2019 (see Figure 11), where unstructured data becomes more important when incorporating the pandemic period into the sample, which is in line with the findings of \cite{richardson2018nowcasting} and \cite{woloszko2020weekly}. Additionally, a contemporaneous correlation analysis of these variables against the monthly GDP is also performed, obtaining that more than 50\% of the unstructured sample correlates greater than 30\%.

The data frequency ranges from daily to monthly records in constructing the model. Each variable was assessed in terms of its predictive ability regarding monthly GDP growth. Then, to facilitate comparison and analysis, we transformed these variables into annualized monthly percentage changes and standardized them. This standardization process allows us to maintain a common reference framework and ensure that different variables contribute equitably to the model.

Ultimately, we have a total set of 91 predictors spanning from January 2008 to May 2023. The evaluation and selection of optimal predictors will be conducted independently for each machine learning algorithm employed. This approach will enable us to refine the process of choosing the most efficient prediction model, thereby achieving enhanced performance.

\subsection{Strategy of the forecast evaluation }
The method that will assess the accuracy in the projection of each model will be done through the root mean square error (RMSE), following the equation:

\begin{equation}
    RMSE = \sqrt{\frac{1}{T} \sum_{t=1}^{T}(y_t - \hat{y}_t)^2}
\end{equation}

where $y_t$ represents the observed value of monthly GDP growth, $\hat{y_t}$ is the forecasted value, and $T$ is the total number of projections made. Following this initial assessment of prediction fit, we will employ the method proposed by \cite{diebold1995paring} to determine if the projections generated by each machine learning model significantly differ from the \textit{benchmark} model.

\section{Results}
\label{sec:results}

This section begins by providing a brief description of the database training period, addresses hyperparameter optimization, and finishes with a thorough analysis of the results. 

\subsection{Estimation and hyperparameters calibration}

To estimate machine learning models, the selection of hyperparameters plays a crucial role in terms of efficiency and accuracy. The optimal determination of these values requires the split of the sample data into three parts: i) a training set, ii) a validation set, and iii) a testing set. Initially, the model is estimated with the training set (in-sample) which turns out the first set of hyperparameters. Then, the search process of the optimal values that minimize the mean quadratic error of projections (\textit{MSE}\footnote{Indicator that measures the average of the squared errors between the predictions of a model and the real values, without applying the square root, used for validation of parameters in ML models.}), through cross-validation techniques. Once it is identified the optimal values, the accuracy of the model is evaluated in the testing set 
(\textit{out-sample}).

\begin{table}[h]
\centering
\caption{Strategy of testing estimations}\label{tab1}
\begin{tabular}{cccccc}
\hline

\hline \multicolumn{5}{c}{\textbf{Training dataset }}& \multicolumn{1}{c}{\textbf{Testing set }} \\ 

 \multicolumn{5}{c}{2008m1-2014m08}& \multicolumn{1}{c}{2014m09-2023m5} \\ \hline

\hline \multicolumn{5}{c}{\textbf{$\longleftrightarrow$}} & \multicolumn{1}{c}{\textbf{$\leftrightarrow$}}\\

Fold 1 & Fold 2 & Fold 3
& Fold 4 & Fold 5 &  \\
\hline
\multicolumn{6}{p{12cm}}{\small Source: Own elaboration}
\end{tabular}
\end{table}

\begin{table}[h]
\centering
\caption{Priors and hyperparameter ranges}\label{tab2}
\begin{tabular}{lccc}
\hline  \multicolumn{1}{c}{\textbf{Model}} & \multicolumn{1}{c}{\textbf{Hyperparameter}} & \multicolumn{1}{c}{\textbf{Range}} & \textbf{Optimised Value} \\ \hline 

\multirow{1}{3cm}{Lasso} & Lambda & 0.001 to 0.009 & 0.007 \\ 
\hline
\multirow{1}{3cm}{Ridge} & Lambda & 0.01 to 0.09 & 0.310 \\ 
\hline
\multirow{2}{3cm}{Elastic Net} & Alpha & 0.1 to 0.9 & 0.500 \\ & Lambda & 0.01 to 0.09 & 0.040 \\
\hline
\multirow{2}{3cm}{Adaptive Lasso} & Lambda & 0.01 to 0.09 & 0.670 \\ & Omega & 0.1 to 0.9 & 0.340 \\ 
\hline
\multirow{1}{3cm}{Random Forest} & \#árboles & 1 to 400 & 281 \\
\hline
\multirow{3}{3cm}{Gradient Boosting Machine} & \# árboles & 1 to 5000 & 19 \\ & Distribución & Normal & Bernoulli \\ & Shrinkage & 0.001 to 0.009 & 0.300 \\ 
\hline
\multicolumn{4}{p{5cm}}{\small 
Source: Own elaboration}
\end{tabular}
\end{table}

The cross-validation method is used to calculate the best hyperparameters with the validation set. This process involves training and validation of the ML model in 5 folds, by using every partition or fold as the validation set and the others as the training set on each iteration. Hence, we obtain 5 performance metrics, one by each fold, which are averaged. Also, to identify the optimal hyperparameters, we will run the cross-validation Bayesian optimization algorithm, following closely \cite{snoek2012practical}.

In addition, to prevent overfitting in the ML models the hyperparameters are bounded within ranges recommended by the reviewed literature (See \cite{zou2005regularization}). This approach contributes significantly to the model's ability to make robust predictions, allowing for more effective exploration in estimating monthly GDP growth without the risk of overfitting. 

\subsection{Model comparison}

A comparison of the prediction performance of the ML and benchmark models for the test set from September 2009 to May 2023 is presented in Table 3.

\begin{table}[!htp]
\centering
\caption{Evaluation of model and benchmark forecasts\\ 2014m09-2023m05}\label{tab3}
\begin{tabular}{lccc}
\hline \multicolumn{1}{c}{\textbf{Model}} & \multicolumn{1}{c}{\textbf{RMSE}} & \multicolumn{1}{c}
{\textbf{RMSE-R.AR\footnote{}}} & \textbf{\emph{p}-value} \\ \hline 
Lasso & 0.26 & 0.10 & 0.014 \\ 
Ridge & 0.34 & 0.13 & 0.043 \\
Elastic Net & 0.28 & 0.11 & 0.039 \\ 
Adaptive Lasso & 0.68 & 0.27 & 0.126 \\ 
Random Forest & 0.45 & 0.18 & 0.089 \\ 
Gradient Boosting Machine & 0.17 & 0.07 & 0.016 \\
\small DFM full  \footnote{} & 0.93 & 0.36 & 0.005 \\ 
\small  DFM best \footnote{} & 0.72 & 0.28 & 0.004 \\ 
\small  DFM structured \footnote{} & 1.05 & 0.41 & 0.003 \\ 
AR & 2.55 & 0.00 &  \\ \hline
\multicolumn{4}{p{13cm}}{\footnotesize Source: Own elaboration. 3/ $RMSE(Model)_i/RMSE(AR)$. 4/ DFM full, use the 91 variables within unstructured as well as structured data. 5/ DFM best, use the 19 variables within unstructured data as well as structured selected by the Gibbs sampling as best estimators to predict GDP. 6/ DFM structured, use only 56 structured variables.} \\
\end{tabular}\\
\end{table}

In terms of the forecast evaluation using the RMSE, the ML models manage to significantly minimize the projection error in comparison with the benchmark AR model and the three different specifications of dynamic factor models \footnote{\cite{banbura2014maximum}}. Between the models that outstand over the others, we get the \textit{Gradient Boosting Machine}, \textit{LASSO} and \textit{Elastic Net} that archive to reduce the forecast error by around 20\% to 25\%. Also, Diebold-Mariano statistic \footnote{\cite{diebold1995paring}} concludes that most of the ML models are statistically significant, in line with previous research. \cite{richardson2018nowcasting, varian2014machine, zhang2023nowcasting}.

On the other hand, it is important to highlight that real-time forecasts presented in this document successfully anticipated the economic contraction caused by the COVID-19 pandemic in March 2020 in the Peruvian context, and also accurately captured the subsequent economic recovery period in March of the following year, which supports the usefulness and effectiveness of using penalty models and/or decision trees to forecast economic variables.

\begin{figure}[!htp]
\centering
\caption{{ML model projection and GDP}}
\noindent 
\begin{minipage}[c]{.5\textwidth}
  \centering
  \includegraphics[width=1\linewidth]{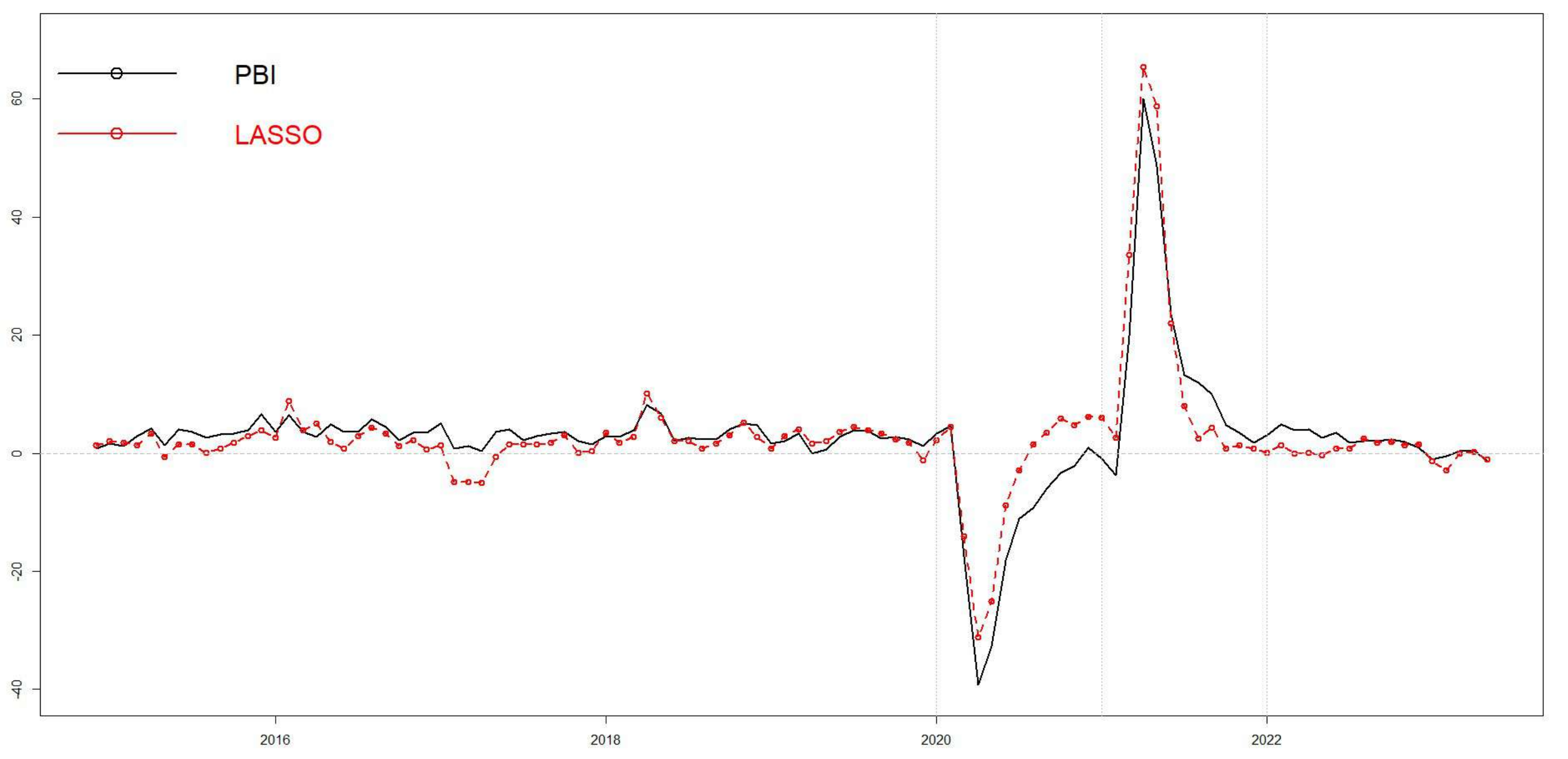}
  \subcaption{LASSO and GDP}\label{fig:1a}
\end{minipage}%
\begin{minipage}[c]{.5\textwidth}
  \centering
  \includegraphics[width=1\linewidth]{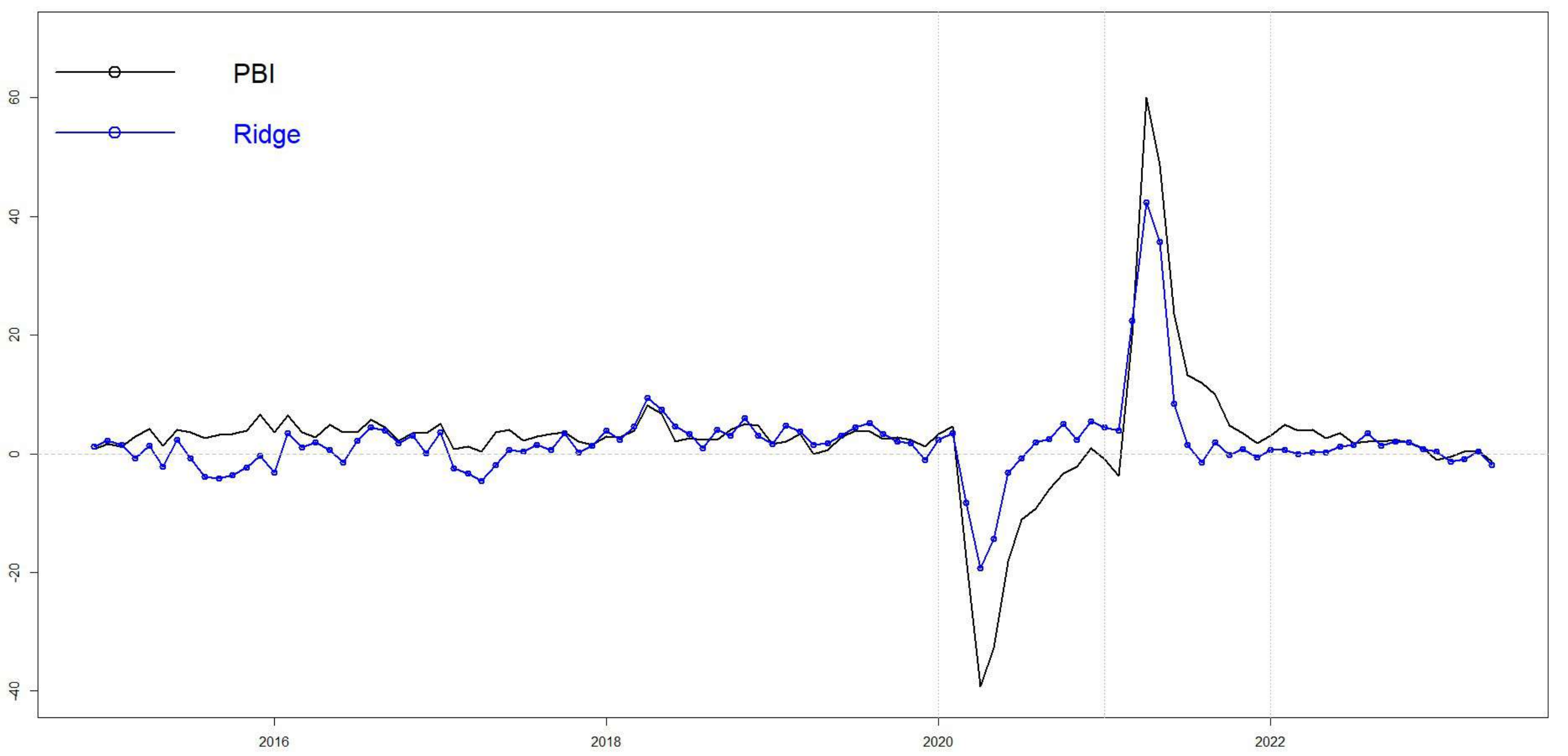}
  \subcaption{Ridge and GDP}\label{fig:2a}
\end{minipage}
\begin{minipage}[c]{.5\textwidth}
  \centering
  \includegraphics[width=1\linewidth]{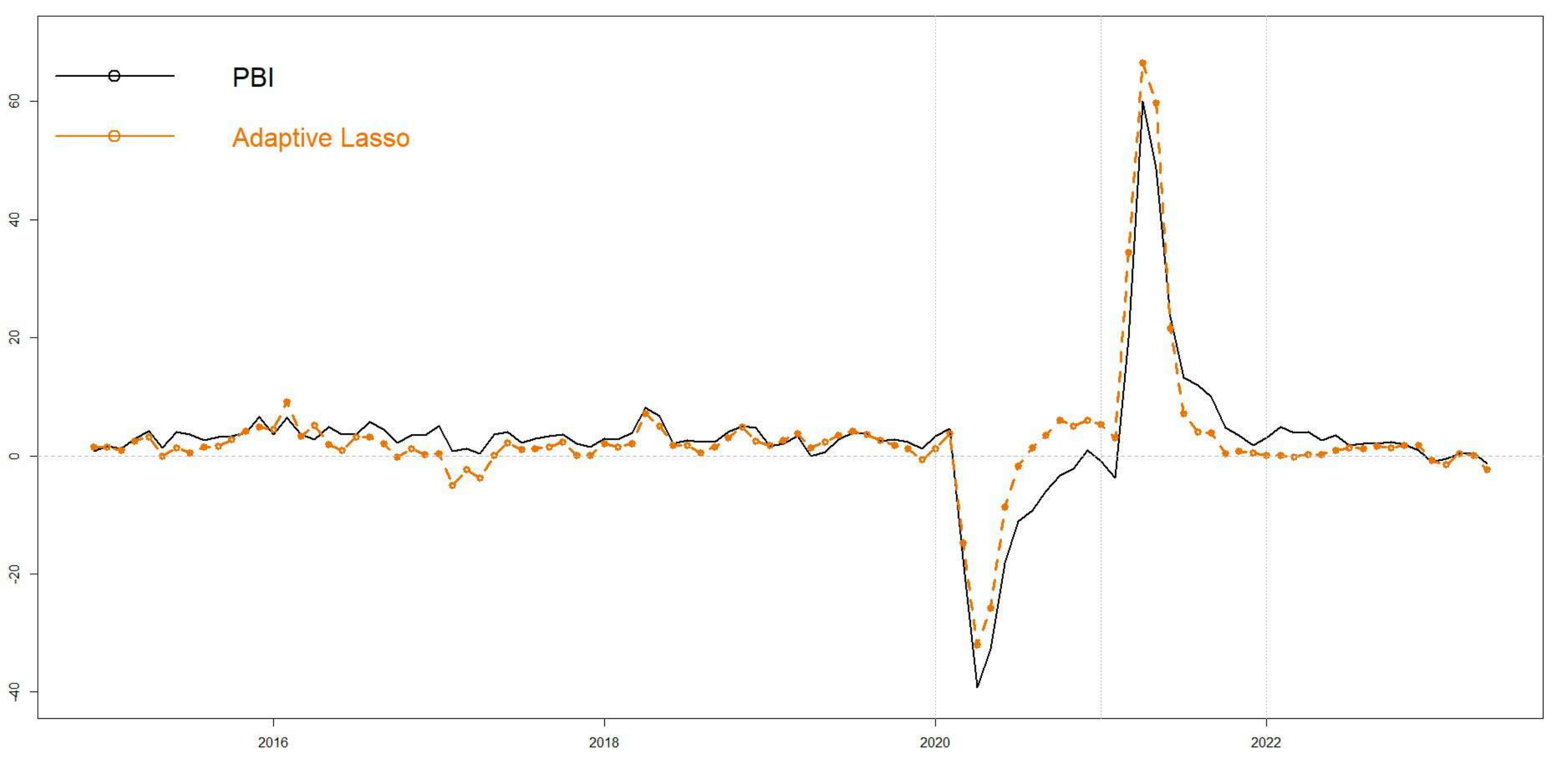}
  \subcaption{Adaptive LASSO and GDP}\label{fig:3a}
\end{minipage}%
\begin{minipage}[c]{.5\textwidth}
  \centering
  \includegraphics[width=1\linewidth]{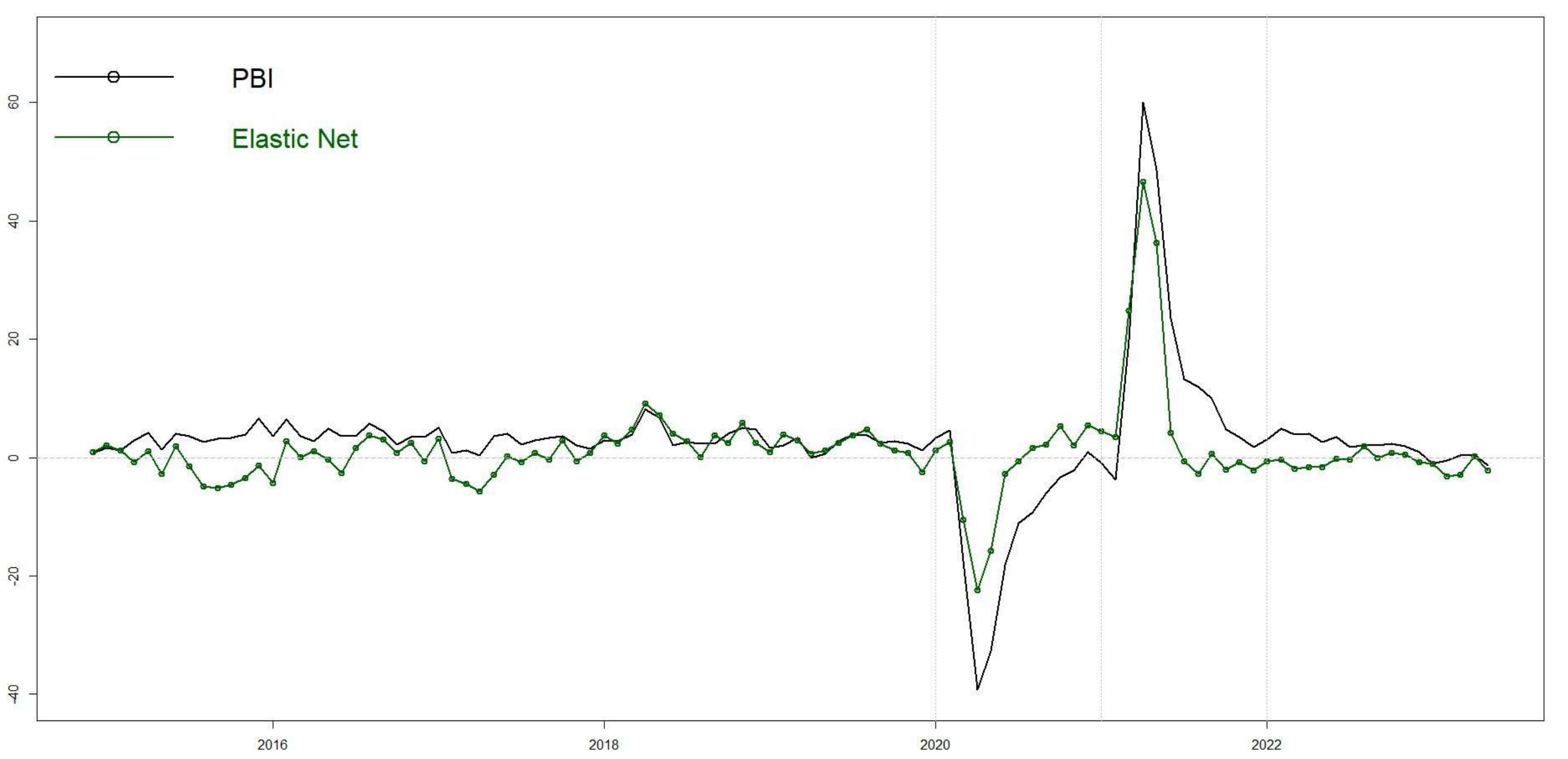}
  \subcaption{Elastic Net and GDP}\label{fig:4a}
\end{minipage}
\begin{minipage}[c]{.5\textwidth}
  \centering
  \includegraphics[width=1\linewidth]{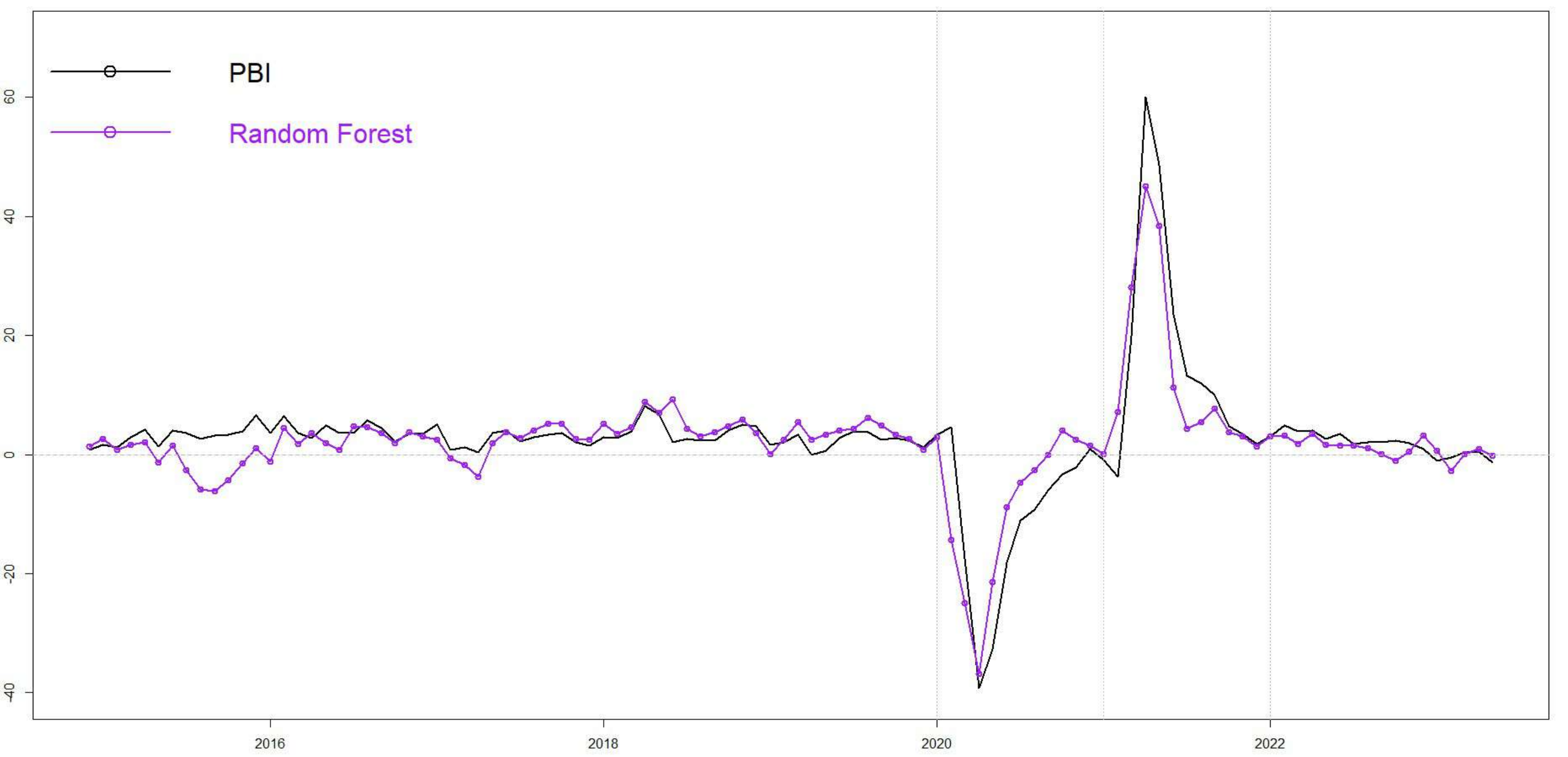}
  \subcaption{Random Forest and GDP}\label{fig:5a}
\end{minipage}%
\begin{minipage}[c]{.5\textwidth}
  \centering
  \includegraphics[width=1\linewidth]{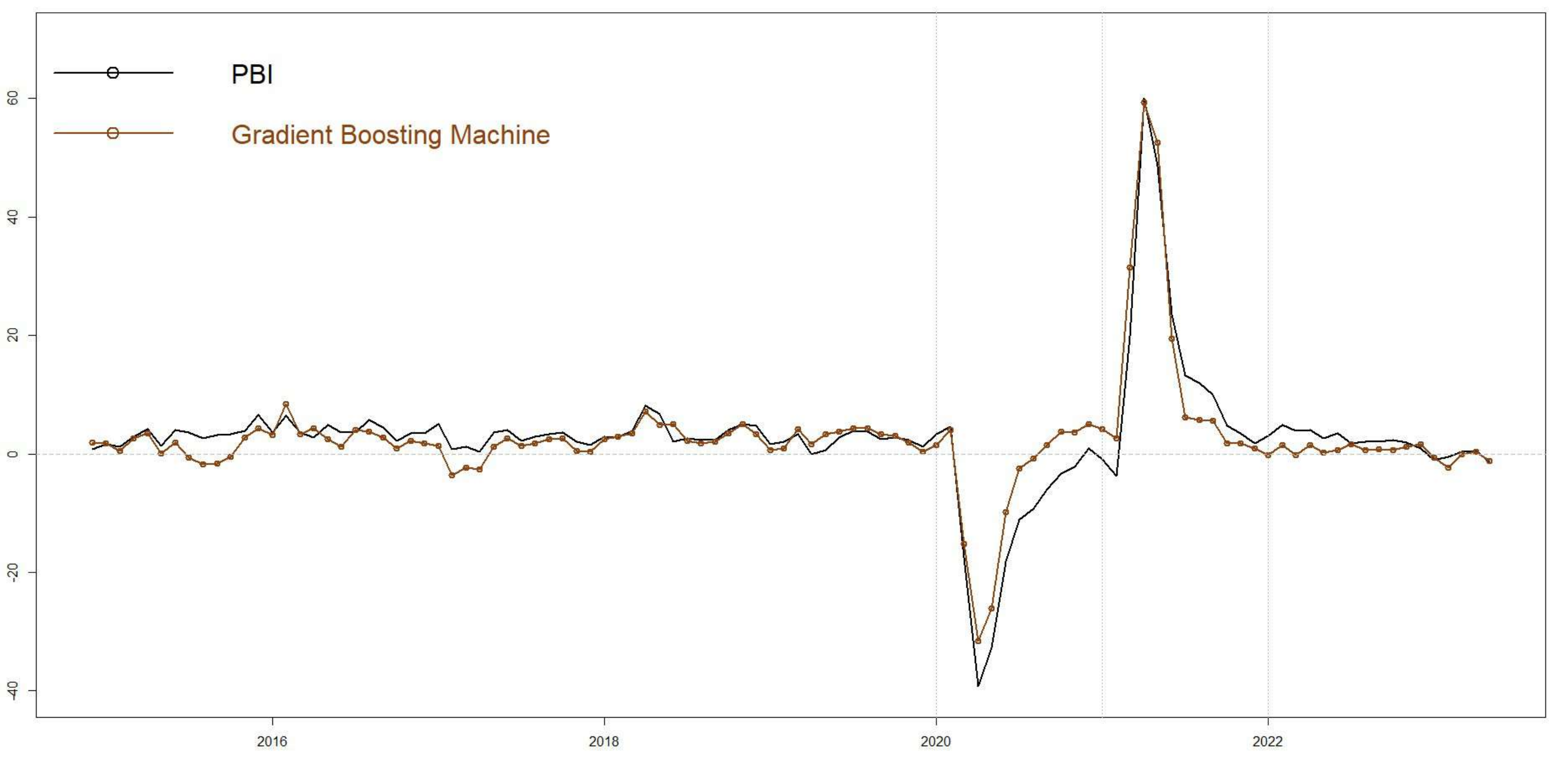}
  \subcaption{Gradient Boosting Machine and GDP}\label{fig:6a}
\end{minipage}
\raggedright
\footnotesize{Source: Own elaboration}
\end{figure}

\subsection{Consistency}
To test the consistency of the results and determine if the ML model projections contribute positively to the accuracy predictions of monthly GDP over the benchmark models, we use the \cite{romer2008fomc} approach, but instead of using an officials prediction, we replace to a DFM estimation that incorporates the electricity as main leading indicator, which popular among Economic Studies Department in Peru. We estimate the following regression model: 

\begin{equation}
    y_t = \beta_1 DFME_t + \beta_2 ML_{it} + e_t
\end{equation}

Where $y_t$ represents the real monthly GDP growth, $DFME_t$ is the dynamic factor model estimated with electricity production and $ML_i$ is the out-sample prediction for each machine learning model. The results obtained indicate that all the projections of machine learning contribute significantly to the GDP projection, with the best model being the \textit{Gradient Boosting Machine} according to the Akaike criterion. Likewise, analyzing the estimation errors of the models generated by equation 14, we applied the test proposed by \cite{harvey1997testing} with a long run variance autocorrelation estimator from \cite{diebold1995paring}, to evaluate the fit gains in the estimates by including the results of the ML models. The $p-value$ is shown in the last column of Table 4, where the alternative hypothesis is that the models in equation 14, which include the ML model projection, are more accurate than the predictions under the dynamic factor model alone. These values indicate a superior accuracy of the models incorporating Machine Learning at a 10\% confidence level in the case of but at 5\% in the others.  

\begin{table}[h]
\centering
\caption{$\beta_2^{e}$ value and validation criteria}\label{tab4}
\begin{tabular}{lcccc}
\hline
\multicolumn{1}{c}{Models} & Estimated & AIC    & $p$-value & \multicolumn{1}{l}{$p$-value (DM)} \\ \hline
Lasso                      & 0.714           & 520.32 & 0.000     & 0.079                              \\
Ridge                      & 0.936           & 554.73 & 0.000     & 0.057                              \\
Elastic Net                & 0.839           & 549.80 & 0.000     & 0.055                              \\
Adaptive Lasso             & 0.703           & 517.49 & 0.000     & 0.046                              \\
Random Forest              & 0.783           & 534.20 & 0.000     & 0.049                              \\
Gradient Boosting Machine  & 0.810           & 492.09 & 0.000     & 0.041                              \\ \hline
{\small 
Source: Own elaboration}
\end{tabular}
\end{table}

\section{Conclusions}
\label{sec:conc}

In this article, we evaluated the prediction accuracy of the most popular Machine Learning algorithms to do the nowcasting of the monthly growth rate of Peruvian GDP. The analysis window was between 2008 and 2023 and works with several leading indicators to assess the dynamic of the GDP's components measured by the expenditure and productive sector approach. Furthermore, it is worth mentioning that we have enriched our approach by incorporating a sentiment data index built through Google Trends, that shown to be helpful to predict in advance economic activity. The Machine Learning approach allows the use of 91 variables simultaneously between structured data and no structured data, one of the documents that use a larger dataset used for the Peruvian GDP prediction case. The evaluation results and consistency exercise show evidence of the positive contribution of ML models and sentiment data improve significantly the model accuracy and allow the early detection of periods of high volatility, an aspect that conventional models often fail to capture.

Our results shed light on outperforming the machine learning over the AR and DFM models in prediction accuracy, which opens a new approach agenda on improvements in the forecast of relevant macroeconomic variables such as consumption, employment, and investment, among others.

In fact, as a pending agenda regard, we can point out three issues. First, there is a need to analyze the marginal prediction gains from the inclusion of unstructured data in reducing forecast error.  Since our results have shown improvements in the accuracy. One question arises. Would the analyzed period between influence those results given that between 2004 and 2023 includes high volatility events such as the pandemic, the global financial crisis and various climate shocks in 2017 and 2023, where ML models with data do not structured ones gain greater predictive capacity by being able to track daily frequency data from Google Trend searches. This could be achieved by performing a variance analysis of the projection errors comparing ML models with other more traditional ones during a period of relative normality and other periods of crisis. Second, a fact we observed in the estimates of the unsynchronized availability of the variables (91), which represented challenges and difficulties, which raises the question of whether consistent results are equally obtained with a smaller number of variables, we estimate this in roughly 45\% of the 91 variables of the dataset.  This proportion could be evaluated in subsequent studies reducing the software requirements. Third, the treatment of the unstructured data could be improved. In this document, we use a simple and didactic management of no-structured data, but it might be considered monthly weighting of searched words in GoogleTrend to smooth the high variability related to this type of data.

\vspace{3cm}

\appendix

\section{List of non-structured and structured variables of model}
\label{sec:sample:appendixA}

\begin{table}[!ht]
\centering
\caption{List of non-structured variables}
\begin{tabular}{ccc}
\hline \multicolumn{3}{c}{\textbf{Unstructured variable details}} \\ \hline

\textbf{Units of Measure} & \textbf{Frequency} & \textbf{Source} \\ \hline

 {Search Index (0 to 100)} & {Daily} & {Google Trends} \\ \hline

\multicolumn{3}{c}{\textbf{Variables}} \\

\hline  \multicolumn{3}{l}{\textbf{1.- Searched Words on Economic}} \\ \hline
\raggedright
 Inflación &  Recesión  &   \\ 

  \hline \multicolumn{3}{l}{\textbf{2.- Searched Words on Consumption}} \\ \hline
 kia &  toyota & Cinema  \\ 
 Restaurantes & Créditos & Préstamos \\
 Hipotecarios & Ofertas & \\ 
\hline \multicolumn{3}{l}{\textbf{3.- Searched Words on Labor Market}} \\ \hline
 Empleo &  Desempleo & Trabajo  \\ 
\hline \multicolumn{3}{l}{\textbf{4.- Searched Words on Sectorial Industry}} \\ \hline
 Minería &  Inversión &   \\ 

\hline \multicolumn{3}{l}{\textbf{5.- Searched Words on Current Situation}} \\ \hline
 Crisis Perú &  Quiebra & Economía  \\ 
 Crisis económica & &  \\ 
\hline
\multicolumn{2}{p{8cm}}{\small 
Source: Own elaboration}
\end{tabular}
\end{table}

\begin{longtable}{lllc}
\caption{List of structured variables included}
\label{tab:structured-variables}\\
\hline 
\multicolumn{1}{c}{\textbf{No.}} & \multicolumn{1}{c}{\textbf{Variable}} & \textbf{Units} & \textbf{Frequency} \\ \hline
\endfirsthead

\multicolumn{4}{c}{{\tablename} \thetable{} -- Continued} \\
\hline 
\multicolumn{1}{c}{\textbf{No.}} & \multicolumn{1}{c}{\textbf{Variable}} & \textbf{Units} & \textbf{Frequency} \\ \hline
\endhead

\hline 
\multicolumn{4}{r}{{Continued on next page}} \\
\endfoot

\hline \hline
\multicolumn{4}{p{11cm}}{\small Note: All variables are included in the model in the form of annualized percentage variation} \\
\multicolumn{4}{p{11cm}}{\small Source: Own elaboration} \\
\endlastfoot

1 & GDP & 2007=100 & Monthly \\ 
2 & Credit & S/ Millions  & Monthly \\ 
3 & Credit & US\$ Millions & Monthly \\ 
4 & Credit (constant exchange rate) & S/ Millions & Monthly \\ 
5 & Consumer credits & S/ Millions & Monthly \\ 
6 & Mortgage Loans & S/ Millions & Monthly \\ 
7 & Deposits  & S/ Millions & Monthly \\
8 & Deposits & S/ Millions & Monthly \\
9 & Sales of chickens & Metric Tons & Daily \\
10 & Consumer Confidence Index & Points & Monthly \\
11 & Electricity Production & GWh  & Monthly \\ 
12 & Hydrocarbon Production & Millios & Daily \\ 
13 & 3-Month Economic Expectations & Points & Monthly \\
14 & Oil & B/D & Daily \\ 
15 & Natural Gas & MCF & Daily \\ 
16 & Domestic Cement Consumption & Index & Weekly \\ 
17 & Import of Intermediate Inputs & Index & Weekly \\ 
18 & Import of Capital Goods & Index & Weekly \\
19 & Employed Labor Force & Thousands & Monthly \\ 
20 & Properly Employed Population \footnote{Metropolitan Lima} & Thousands & Monthly \\
21 & Non-Financial Gov. Expenditures & S/ Millions & Monthly \\ 
22 & IAFO & Index & Monthly \\ 
23 & Volume of Imported Inputs & Index & Monthly \\ 
24 & Terms of Trade  & Index & Monthly \\ 
25 & IPX & Index & Monthly \\
26 & IPM & Index & Monthly \\
27 & General Stock Market Index\footnote{Lima} & Percentages & Daily \\ 
28 & Liquidity & Millions of Soles & Monthly \\ 
29 & CPI & Index & Monthly \\ 
30 & Non Food and Energy Price Index & Index & Monthly \\ 
31 & Wholesale Price Index & Index & Monthly \\
32 & Core CPI & Index & Monthly \\
33 & Multilateral Real Exchange Rate & 2009=100 & Monthly \\ 
34 & EMBIG Peru & Pbs & Daily \\ 
35 & Oil WTI & Dollars per Barrel & Daily \\
36 & USIPC & Index & Monthly \\
37 & Industrial Production Index & YoY & Quarterly \\ 
38 & Copper & cUS\$/lb. & Daily \\ 
39 & Gold & US\$/oz.tr. & Daily \\
40 & US Manufacturing PMI & Points & Monthly \\
41 & FED Interest Rate (Upper Limit) & Percentages & Monthly \\ 
42 & VIX Index & Percentages & Daily \\ 
43 & Spread 2Y-5Y & Points & Monthly \\
44 & China Industrial Production & YoY & Monthly \\
45 & PPI by All Commodities & 1982=100 & Monthly \\
46 & ATSM & Degrees Celsius & Monthly \\ 
47 & Anchovy Landing & Metric Tons & Daily \\ 
48 & Logarithm of Anchovy Landing & Index & Daily \\ 
49 & Anchovy Landing\footnote{Seasonally Adjusted} & Index & Daily \\ 
50 & Variation Anchovy Landing\footnote{Seasonally Adjusted} & YoY & Daily \\
51 & Paddy Rice production & Tons & Monthly \\ 
52 & Potato production & Tons & Monthly \\  
53 & Onion production & Tons & Monthly \\ 
54 & Tomato production & Tons & Monthly \\
\end{longtable}

\newpage
\section{Gibb sampling estimations}
\label{sec:sample:appendixB}

\begin{figure}[!htp]
\centering
\caption{{Gibb sampling (2004-2023) - probability of inclusion in optimal model}}
\noindent 
\begin{minipage}[c]{1\textwidth}
  \centering
  \includegraphics[width=1\linewidth]{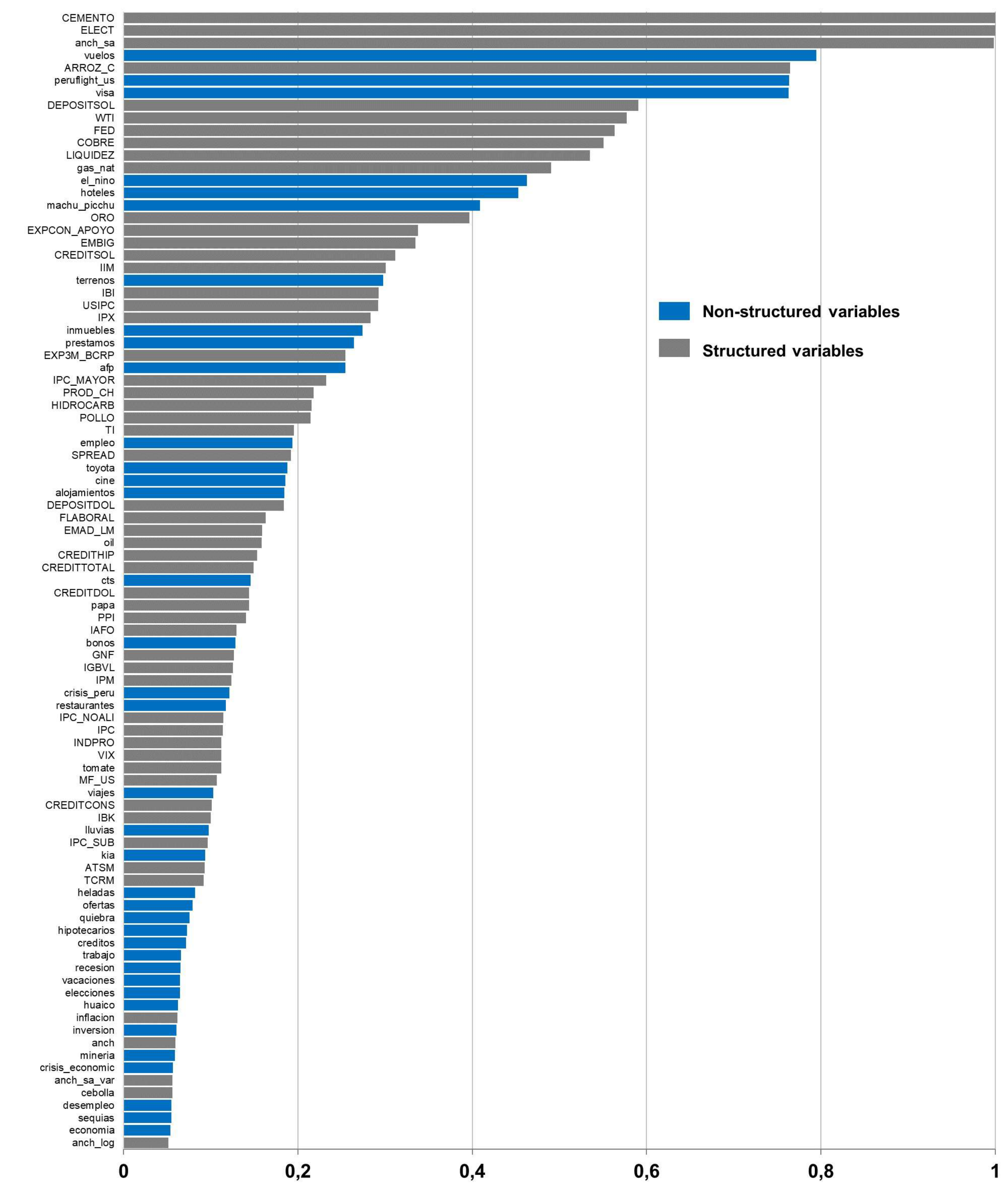}
\end{minipage}%
\end{figure}

\begin{figure}[!htp]
\centering
\caption{{Gibb sampling (2004-2019) - probability of inclusion in optimal model}}
\noindent 
\begin{minipage}[c]{1\textwidth}
  \centering
  \includegraphics[width=1\linewidth]{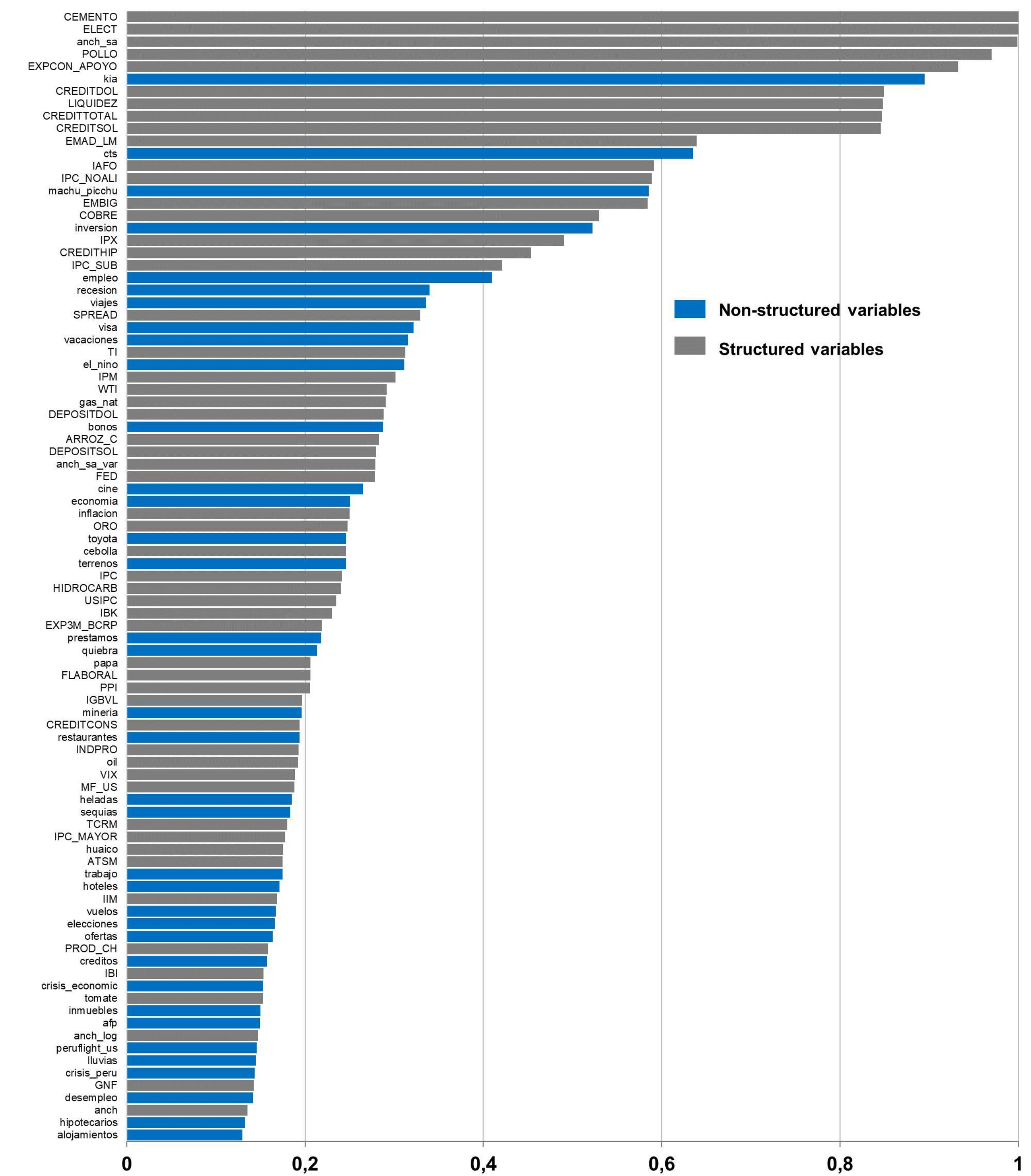}
\end{minipage}%
\end{figure}

\newpage

 \bibliographystyle{elsarticle-num} 
 \bibliography{cas-refs}





\end{document}